# Grain boundary-mediated nanopores in molybdenum disulfide grown by chemical vapor deposition

Kenan Elibol[1], Toma Susi[1], Maria O'Brien[2], Bernhard C. Bayer[1], Timothy J. Pennycook[1], Niall McEvoy[2], Georg S. Duesberg[2], Jannik C. Meyer[1], Jani Kotakoski[1,*]

[1]Faculty of Physics, University of Vienna, Boltzmanngasse 5, 1090 Wien, Austria

[2]School of Chemistry, Centre for the Research on Adaptive Nanostructures and Nanodevices (CRANN) and Advanced Materials and BioEngineering Research (AMBER), Trinity College Dublin, Dublin 2, Ireland

*Email: jani.kotakoski@univie.ac.at

**Molybdenum disulfide ($MoS_2$) is a particularly interesting member of the family of two-dimensional (2D) materials due to its semiconducting and tunable electronic properties. Currently, the most reliable method for obtaining high-quality industrial scale amounts of 2D materials is chemical vapor deposition (CVD), which results in polycrystalline samples. As grain boundaries (GBs) are intrinsic defect lines within CVD-grown 2D materials, their atomic structure is of paramount importance. Here, through atomic-scale analysis of micrometer-long GBs, we show that covalently bound boundaries in 2D $MoS_2$ tend to be decorated by nanopores. Such boundaries occur when differently oriented $MoS_2$ grains merge during growth, whereas the overlap of grains leads to boundaries with bilayer areas. Our results suggest that the nanopore formation is related to stress release in areas with a high concentration of dislocation cores at the grain boundaries, and that the interlayer interaction leads to intrinsic rippling at the overlap regions. This provides insights for the controlled fabrication of large-scale $MoS_2$ samples with desired structural properties for applications.**

# Introduction

Two-dimensional (2D) materials such as graphene, hexagonal boron nitride (hBN), and transition metal dichalcogenides like molybdenum disulfide ($MoS_2$) show great potential for applications, especially in their monolayer (ML) forms (for examples, see Ref. 1). A monolayer of $MoS_2$ consists of two atomic layers of sulfide (S) and one atomic layer of molybdenum (Mo) between the sulfide layers. Within each layer, the S-Mo-S bonds are covalent, whereas the interlayer interaction is dominated by weak van der Waals forces[2,3]. $MoS_2$ exhibits semiconducting character due to its moderate band gap[4] in contrast to semi-metallic graphene and insulating hBN. The attractive properties of $MoS_2$ originate from its unsaturated $d$-electron interactions and quantum confinement[2,5,6] and can be tuned by applying strain, through defect engineering, or by controlling its rippling[7–9]. Currently, chemical vapor deposition (CVD) is the only way to produce macroscopic samples of ML $MoS_2$. Since CVD leads to the growth of polycrystalline samples, understanding the atomic structure of grain boundaries is important for the growth of high-quality large area $MoS_2$ films suitable for applications.

Here we present an atomic-scale scanning transmission electron microscopy (STEM) analysis of the longest grain boundary (GB) areas reported to date in CVD-grown $MoS_2$. Covalently bound (stitched) GBs in CVD-$MoS_2$ occur via the merging of triangular grains during the growth process[10,11]. When two grains grow together but do not merge, an overlap region with a bilayer (BL) structure is formed instead. Surprisingly, we observed a chain of nanometer-sized pores (nanopores, NP) in all studied stitched GBs with size distributions that appear to be associated with the angle between the two grains: a decreasing misorientation angle leads to an increasing density of dislocation cores (DCs)[12–14] and to larger NPs.

Despite being detrimental to some properties, NPs in 2D materials can also be beneficial. For example, they show promise for DNA sequencing, single molecule detection, filtration and power generation[15,16]. In fact, NPs in $MoS_2$ are thought to be particularly suited for, e.g., water desalination[15,17]. The spontaneous formation of pores during film manufacture may provide an alternative to hitherto used dedicated post-growth processing via electron or ion beam irradiation or by electrochemical reactions[18–21]. In addition to reducing the number of required processing steps, this also avoids drawbacks of those other methods such as material degradation during irradiation and contamination during chemical processes.

Our analysis of the BL areas at overlap regions reveals intrinsic rippling resulting from the Moiré pattern of just a slight misorientation between the grains (ca. 0.8°). Due to the sensitivity of the electronic system of $MoS_2$, this is expected to lead to local variation of the optical and electronic properties due to varying interlayer separation and stacking as well as intralayer strain, revealing another way to affect the properties of CVD-$MoS_2$ directly through the growth process.

**Results and discussion**

Our $MoS_2$ samples were synthesized on $SiO_2$/Si substrates by CVD in a micro-cavity setup that allowed the solid state precursor to be supplied in close proximity[22]. They were transferred onto carbon-supported transmission electron microscopy (TEM) grids (for STEM experiments) and SiN/Si chips with multiple holes (for other experiments) following the method reported in Ref. [23]. Formation of different grain boundary areas is schematically illustrated in Figure 1. CVD growth of $MoS_2$ on $SiO_2$ occurs through the growth of randomly placed and oriented triangle-shaped grains (see the atomic force microscopy image in Supplementary Figure 1) that merge together[24]. The colored frames in Figure 1 show high angle annular dark field (HAADF)-STEM images of $MoS_2$ including a stitched GB between two ML grains (red), ML without a grain boundary (yellow) and two overlapping regions corresponding to BL stacking with a 0.8° (blue) and a 10° (green) rotation between the layers. While a clear Moiré pattern occurs for the larger rotation angle, it is not apparent for the structure with the smaller rotation due to its large Moiré period. This schematic also shows how the grains create triangular empty spaces when the growth finishes prematurely (see also Supplementary Figure 2).

In Figure 2, we present large-scale images of stitched GBs in CVD-$MoS_2$ along with the corresponding NP size distributions. The insets of Figure 2a-c show HAADF-STEM images of three different samples with different but relatively similar misorientation angles between the grains. We observed NPs at all analyzed stitched GBs, but not within the grains themselves. We point out that although electron irradiation in a TEM experiment can cause nanopore formation in $MoS_2$ even in near ultra-high vacuum due to sputtering[25], such pores are easy to distinguish from the intrinsic ones while imaging (see also the Supplementary Material for an irradiation-induced nanopore and two examples of the evolution of a GB area during imaging). For the GBs shown in Figure 2, the NP areas varied from 0.12 nm$^2$ to 25.45 nm$^2$ with the larger pores being comparatively rare (less than ~5% of pores had areas larger than 10 nm$^2$). Although we could not directly ascertain this, the few rare larger pores may have been formed during the sample transfer process. If this is the case, these might be avoided by improving the sample preparation method. The geometric mean sizes (± standard error) of the pores are 0.82 ± 0.17, 0.74 ± 0.20 and 0.60 ± 0.13 nm$^2$ for misorientations of 7.9°, 8.9° and 9.6°, respectively. These misorientation angles were measured directly from the STEM images for atomic rows next to the boundaries with a field of view of 20 nm. Misorientations measured from Fourier transforms of the same images (dominated by atomic rows away from the GB) lead to values of 8.7°, 9.2° and 10.5°, respectively. The difference stems from local adjustment of the grain orientations toward each other at the boundary. Notice that the first bins in the histograms (smallest pores in inset of Figure 2a-c) correspond to single and double Mo vacancies along with their neighboring S atoms. The longest atomic-scale images of MoS$_2$-GBs, created by merging several HAADF-STEM images, are shown in Figure 2d-f (see Supplementary Material for full-resolution atomically resolved composite images of the GBs). The bright contrast along the GBs is caused by hydrocarbon contamination attracted to the chemically more reactive GB structures (as compared to pristine MoS$_2$). We point out that this contamination is also present in most of the NPs observed during STEM imaging, although its contrast in the images is lower than that of MoS$_2$.

To further understand the formation of GB-NPs, we analyzed the dislocation cores along the stitched GBs, as shown in Figure 3. The original HAADF-STEM image (in false color) is presented in Figure 3a. In Figure 3b, the same image is treated with a Wiener filter to enhance the contrast of the atomic structure. On top of this image, we have indicated the locations of five easily visible DCs as well as the NP by drawing Burger circuits around them. The location of each DC is confirmed by Figure 3c, which shows a strain map obtained from geometric phase analysis (GPA)[26–28] of the same image (DCs are marked with squares): the highest amounts of tensile and compressive strains appear at the DCs. Interestingly, it can be seen that one DC (highlighted by the white square) was hidden under the bright contamination directly left of the NP. A more detailed analysis of the strain around one of the DC structures is presented in Figure 3d-f. As can be seen from the plots, the $\varepsilon_{xx}$ strain fits both to the Peierls-Nabarro (PN) and the Foreman (FM) dislocation models[29]. Indeed, for the FM model, the experimental strain profile is best fitted with a parameter $a = 1$ (see Supplementary Material), at which point the two models converge. We point out that although it is difficult to measure absolute strain values with a scanning device due to the possibility of sample drift, its (possible) outcome would be a uniform strain field superimposed on the actual strain, which would not alter any of our conclusions.

We hypothesize that the NP formation is associated with a buildup of stress at GBs with a high concentration of DCs. As can be seen in Figure 3, the NP covers an area in which there would otherwise need to be four DCs. While the surrounding area shows that such a DC density is indeed possible, it is likely that the two grains would not be able to sustain the resulting strain over long distances. Analysis of the DC density of the three GBs presented in Figure 2 is shown in Figure 4. We start again with atomic-resolution images of example areas of the GBs (Figure 4a-c). Figure 4d-f show the corresponding mapping of $\varepsilon_{xx}$, $\varepsilon_{yy}$ and shear ($\varepsilon_{xy}$) strain fields and rotation ($\gamma_{xy}$) obtained from GPA. Again, the points indicating the highest levels of tensile and compressive strain fields are DCs in the GBs. The insets in Figure 4a-c show the distributions of distances between DCs in each GB. The average distance (± standard error) between the dislocation cores is measured as 2.12 ± 0.04, 2.01 ± 0.07 and 1.89 ± 0.03 nm for the 7.9°, 8.9° and 9.6° GBs, respectively. Even with the admittedly relatively low statistics, this clearly indicates that the DC separation decreases for increasing misorientation. To quantify this result, we compare it against the Frank–Bilby equation, given as $\delta = b/[2 \sin(\theta/2)]$ in Ref. [30] (where $b$ is magnitude of the Burger vector and $\delta$ is the spacing of GB DCs for the misorientation angle $\theta$). According to the equation, a large spacing between DCs should be observed for a small misorientation angle, consistent with our results. The values are shown in Table 1 along with our experimental estimates.

**Table 1.** The average distances ($\delta_{exp}$ ± standard error) between dislocations and the geometric mean of the sizes (A ± standard error) of nanopores for each misorientation angle. $\delta_{FBE}$ are the expected dislocation distances based on the Frank-Bilby equation.

| θ (°) | $\delta_{exp}$ (nm) | $\delta_{FBE}$ (nm) | A (nm$^2$) |
|---|---|---|---|
| 9.6 | 1.89 ± 0.03 | 1.86 | 0.60 ± 0.13 |
| 8.9 | 2.01 ± 0.07 | 2.01 | 0.74 ± 0.20 |
| 7.9 | 2.12 ± 0.04 | 2.26 | 0.82 ± 0.17 |

Unfortunately, due to the hydrocarbon contamination, the exact atomic structure at the DCs remain in most cases elusive even in our atomic-resolution images. However, in Supplementary Figure 3, we present an example of a clean DC with its complete atomic structure resolved and compared to a model simulated with density functional theory (DFT).

Finally, we turn to the analysis of the overlap areas. It has already been shown that triangular $MoS_2$ grains can grow on top of each other without creating chemical bonds[10]. In our example case, the rotation between the two $MoS_2$ layers is 0.8° with a Moiré periodicity of ~14 nm, as shown in Figure 5. Our DFT calculations (see Methods) confirm that AB stacking is energetically favored over AA[31] stacking and the AB-AA transition region, all of which are observed in the experimental image. Since these different stackings also have different equilibrium distances (by less than 0.6 Å), it is reasonable to expect that one or both of the overlapping layers have an undulating structure. As a simple model of this situation, we created two model grains with the same misorientation as in the experiment, and artificially introduced a small ripple to one of them so that the different stackings each have an interlayer distance consistent with DFT calculations, and with a smooth transition between them with the same periodicity as in the experimental image. The simulated HAADF-STEM image based on this model (Figure 5b) agrees well with the experiment, indicating that even such a small rotation leads to intrinsic rippling in BL $MoS_2$. This result is also consistent with our analysis of the electron diffraction patterns of BL areas within the samples (see the Supplementary Material). Due to the sensitivity of the electronic properties of $MoS_2$ to the stacking order (see Supplementary Material) and external influences[32], this allows at least in principle tuning those properties by controlling the misorientation of the overlapping grains.

**Conclusions**

Through atomic-resolution analysis of the longest grain boundary structures in $MoS_2$ samples reported to date, we showed that stitched boundaries tend to be decorated by a chain of nanopores. The boundaries that we found formed between grains with misorientation angles ranging from 7.9° to 9.6°, and the observed pores had geometric mean sizes between 0.82 and 0.60 $nm^2$ with a tendency for decreasing pore sizes with increasing angle. Through geometric phase analysis, we show that the average distances between the dislocation cores at $MoS_2$ grain boundaries is greater for larger misorientation angles following the Frank-Bilby equation, which can help estimating the expected pore size for arbitrary misorientation angles. Further, we also presented an atomic-resolution analysis of an overlap region where two $MoS_2$ grains have joined without covalent bonding, resulting instead in the formation of a bilayer with a misorientation of approximately 0.8°. Our analysis shows that the Moiré structure of the two overlapping lattices leads to a rippling of the material due to the different interlayer distances for different $MoS_2$ stackings, hinting towards a possibility to create bilayer structures with locally varying electronic properties. Although defective grain boundaries have certainly a negative impact on properties important for some potential applications, the large number of possible applications for nanoporous $MoS_2$ in sensing, DNA sequencing and water desalination gives promise that careful control over the grain orientation and the grain boundary structures during CVD-growth may lead to significant advances in large scale production of $MoS_2$-based devices.

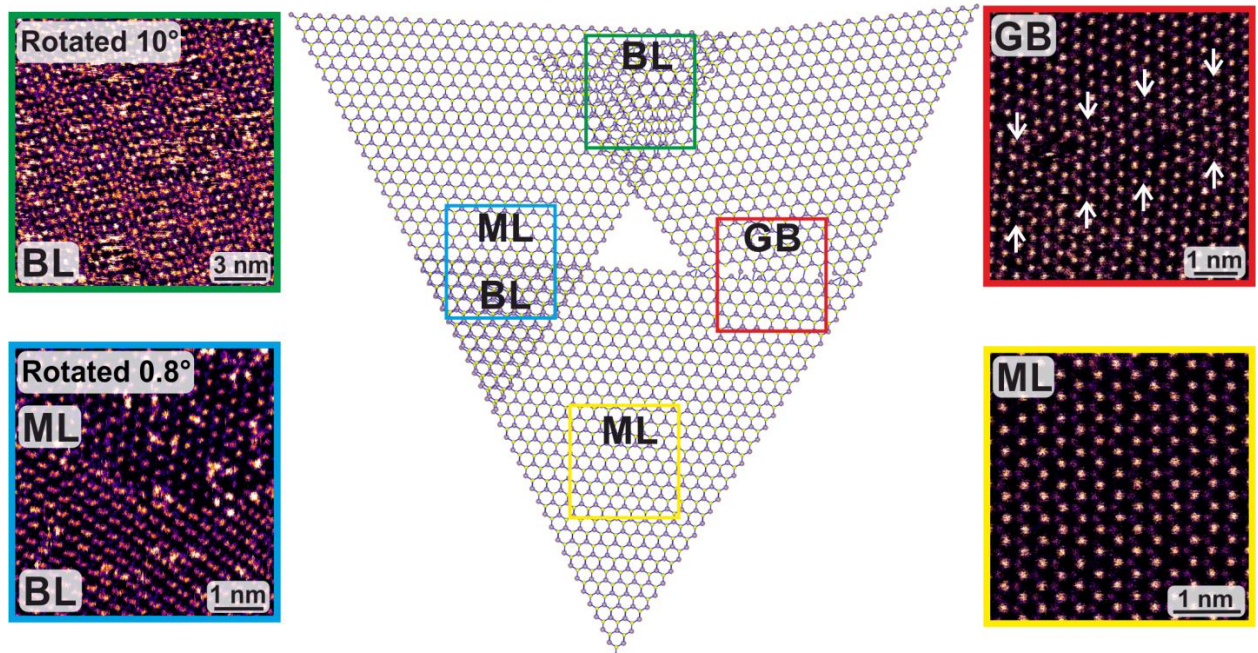

**Figure 1. A schematic representation of the formation of stitched grain boundary (GB) structures between two monolayer (ML) grains and bilayer (BL) stacking in overlap areas.** HAADF-STEM images of MoS$_2$ GB and BL stacking of MoS$_2$ are shown in the colored frames.

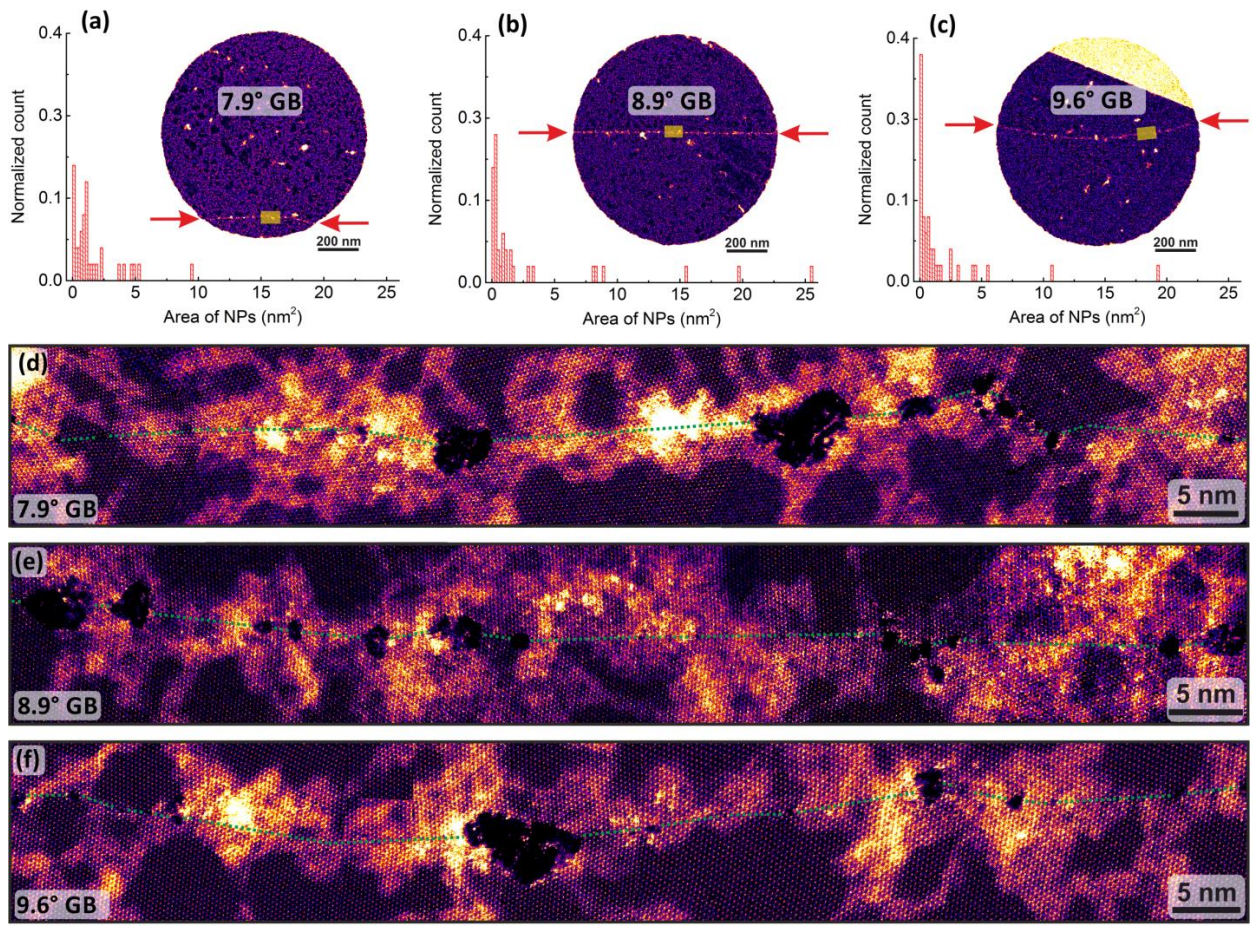

**Figure 2. Large-scale images of stitched grain boundaries.** (a-c) Histograms showing the area distribution of nanopores (NP) for 7.9°, 8.9° and 9.6° angle boundaries. Insets are HAADF-STEM images showing overviews of each of the analyzed structures. The red arrows show the location of the boundaries (they are also easy to recognize from the brighter contrast resulting from hydrocarbon contamination). (d-f) Atomic-resolution HAADF-STEM images of the areas marked with yellow rectangles in the overview images of panels (a-c). The green dashed lines indicate the boundary between the grains.

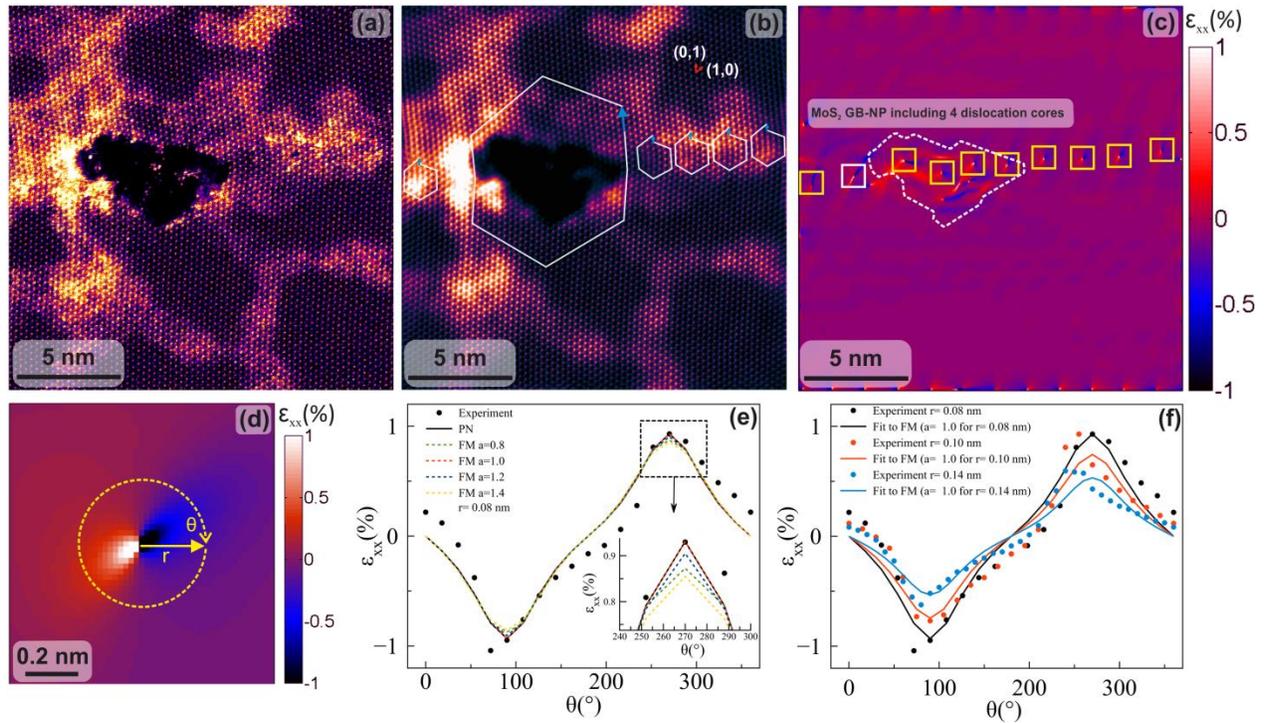

**Figure 3. Stress analysis of a nanopore and dislocation cores at a stitched grain boundary.** (a) HAADF-STEM image of a short section of a boundary with a nanopore. (b) The same image with a Wiener filter applied. White lines indicate Burger circuits around dislocation cores. The red and blue arrows show the lattice unit and Burger vectors, respectively. (c) $\varepsilon_{xx}$ strain map of the image. Dislocation cores are indicated by squares (the white square shows a location of a dislocation core hidden under contamination in panels (a) and (b)). (d) Strain map around one of the dislocation cores. (e) $\varepsilon_{xx}$ strain profile taken at a radius of 0.08 nm for image in panel (d) and fits to two dislocation models (PN and FM) with different fitting parameters. (f) $\varepsilon_{xx}$ strain profiles taken at different radii along with fits to the FM dislocation model with parameter $a = 1$.

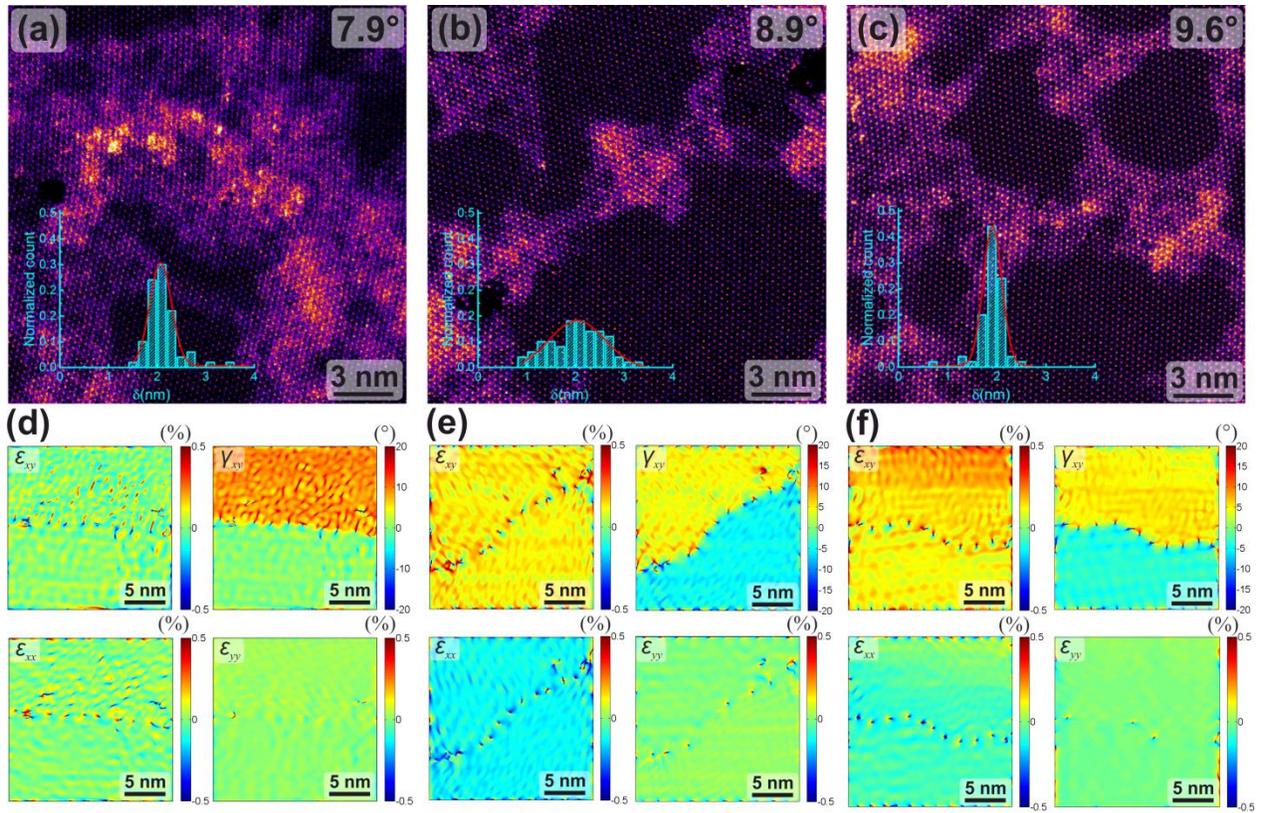

**Figure 4. Dislocation density analysis at stitched grain boundaries.** (a-c) Atomic-resolution HAADF-STEM images of the three grain boundaries. Insets show the distributions for distances between the dislocation cores for each boundary. (d-f) Strain maps for each boundary, calculated with geometric phase analysis: shear strain ($\varepsilon_{xy}$), rotation ($\gamma_{xy}$), $\varepsilon_{xx}$ and $\varepsilon_{yy}$ strain field. The features at the edges of the images are boundary artifacts.

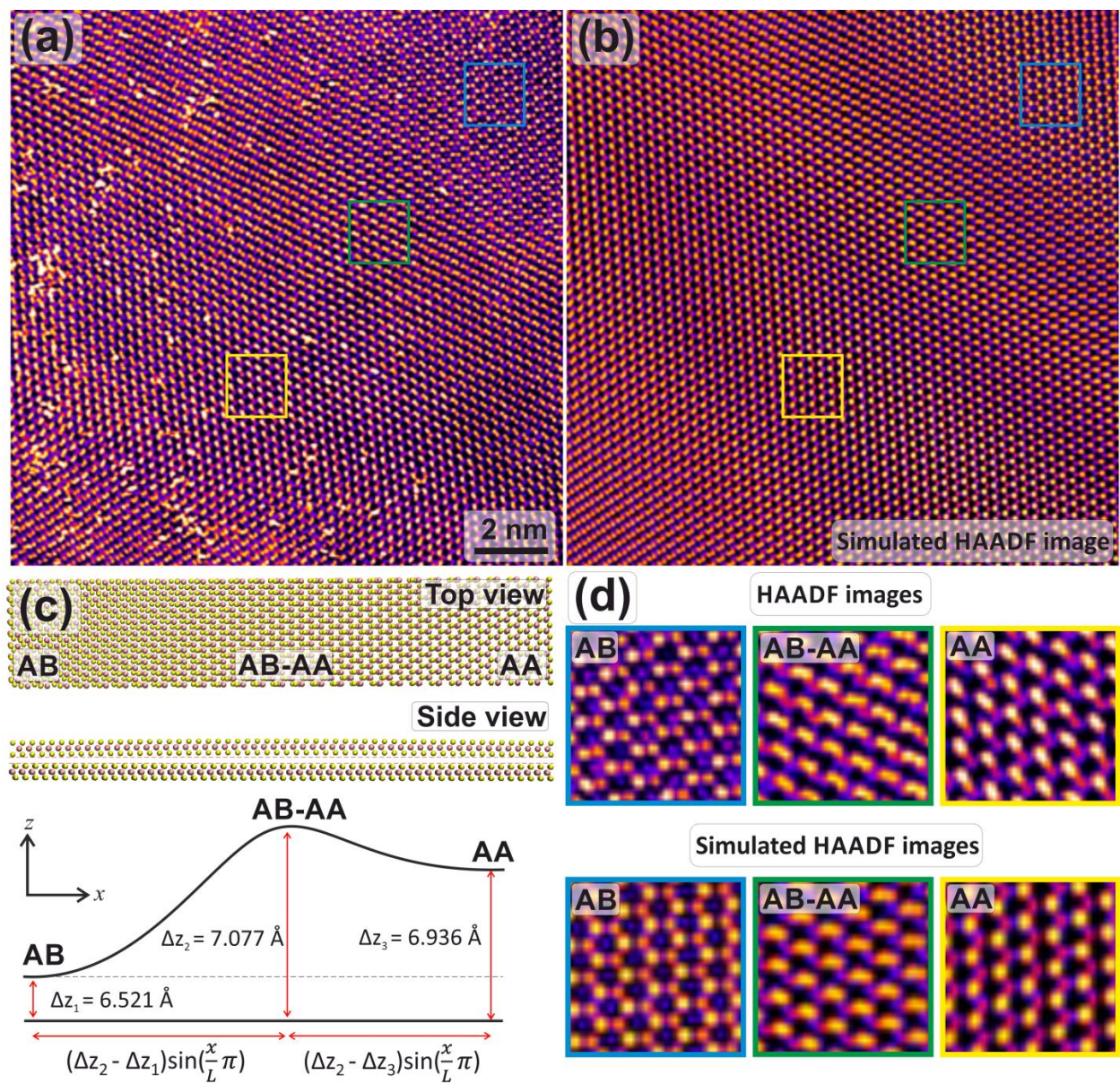

**Figure 5. Atomic-resolution analysis of a bilayer overlap area with a misorientation angle of 0.8°.** (a) Double Gaussian filtered[33] HAADF-STEM image of the area (unprocessed image is included as Supplementary Figure 4). Squares with different colors on the image show different stacking orders. (b) Simulated HAADF image corresponding to (a). (c) The model used in the simulations. Note that the undulation, albeit small, is visible in the side view image. The graph under the model shows the calculated interlayer distances for different stacking orders and the equations describe the smooth transition between them as implemented for our model structure (*L* denotes distance between the different stackings). (d) Experimental and simulated images from the highlighted frames of (a) and (b).

**Methods**

*Sample growth* – The MoS$_2$ samples were synthesized on SiO$_2$/Si using the CVD microreactor method described in Ref. [22] and transferred onto carbon-supported TEM grids or SiN/Si chips with multiple holes following the method reported in Ref. [23].

*Transmission electron microscopy* – STEM images were obtained using a Nion UltraSTEM 100 electron microscope with an accelerating voltage of 60 kV in near ultra-high vacuum (2×10$^{-7}$ Pa) using the high angle annular dark field (HAADF) detector with a collection angle of 80-200 mrad. Diffraction experiments (Supplementary Material) were carried out using a Delong instruments LVEM5 table-top transmission electron microscope operated at 5 kV.

*Modeling* – Our atomistic simulations were carried out with density functional theory as implemented in the GPAW software package[34,35]. To study the energetics of MoS$_2$ bilayers, we created models with two six-atom unit cells of MoS$_2$ in AA, AB and intermediate stacking. Using a plane-wave basis (cutoff energy 700 eV, 16×16×1 *k*-point mesh) we relaxed the atomic positions using the C09 van der Waals functional[36] (we tested other exchange correlation functionals but this one seemed to give the best agreement with Refs. [31,37]) so that maximum forces were below 0.01 eV/Å. Comparing the total energies (*E*) for each stacking order, we found $E_{AB}$, $E_{AA}$, and $E_{AA-AB}$ to be 22.926 eV, 22.971 eV and 22.969 eV, respectively. The interlayer distances for the AA, AB-AA and AB stackings were 6.936 Å, 7.077 Å and 6.521 Å, respectively. These results were used in the model structure (Figure 5). To model the dislocation core structure (Supplementary Figure 3), we created a 273-atom model and relaxed its atomic positions with periodic boundary conditions and a 3×3×1 *k*-point mesh so that maximum forces were <0.05 eV/Å. Since this large structure has 2352 electrons, we used a double-zeta linear combination of atomic orbitals basis to speed up the calculations[38]. The STEM image simulations were carried out with the QSTEM software[39] with parameters matching our experimental conditions.

*Atomic force microscopy and Raman spectroscopy* – Atomic force microscopy (AFM) and Raman spectroscopy results (Supplementary Figure 1) were recorded using a NT-MDT

NTEGRA Spectra combined AFM-Raman spectrometer with a 473 nm wavelength. A Pt coated Si AFM cantilever with visual access (NT-MDT VIT_P_Pt) was used in the AFM experiment.

**Supplementary Material**

Supplementary Material contains AFM and Raman spectroscopy results, STEM-HAADF images of triangular holes between grains, an atomic-resolution image of a dislocation core and the corresponding simulated model, image of a transition between a monolayer and bilayer area, TEM diffraction analysis, strain analysis of an electron-beam-induced nanopore, theoretical equations describing the strain around dislocations, untreated image of the structure shown in Figure 5a and a table of energies, interlayer distances and band gaps calculated for bilayers with different stacking orders. References [40–46] are cited in the Supplementary Material. Supplementary files with atomic-scale images of the three studied covalently bound grain boundaries are additionally provided as full-resolution composite figures.

**Acknowledgments**

K.E. and J.C.M. acknowledge support from the Austrian Science Fund (FWF) through project P25721-N20 and T.S. through P28322-N36. B.C.B. and T.J.P. acknowledge funding from the European Union's Horizon 2020 research and innovation programme under the Marie Skłodowska-Curie grant agreement No. 656214-2DInterFOX (B.C.B.) and 655760-DIGIPHASE (T.J.P.). M.O.B. acknowledges an Irish Research Council scholarship via the Enterprise Partnership Scheme, Project 201517, Award 12508. N. M. acknowledges support from SFI through 15/SIRG/3329. G.S.D. acknowledges the support of SFI under Contract No. 12/RC/2278 and PI_10/IN.1/I3030. J.K. acknowledges support from the Vienna Science and Technology Fund (WWTF) through project MA14-009. Computational resources through the Vienna Scientific Cluster are also gratefully acknowledged.

# Supplementary Information

**Atomic force microscopy and Raman spectroscopy analysis**

A Raman spectrum indicating the in-plane $E^1_{2g}$ and out-of-plane $A_{1g}$ modes of $MoS_2$ is shown in Supplementary Figure 1a. The peak positions of the $E^1_{2g}$ and $A_{1g}$ Raman modes are ca. 382 cm$^{-1}$ and ca. 405 cm$^{-1}$, which confirm[40,41] the formation of BL $MoS_2$ (Ref. [42]). The 2LA(M) Raman peak is also observed at the position of ca. 452 cm$^{-1}$ (Refs. [43,44]). Supplementary Figure 1b shows a tapping mode AFM topography image of $MoS_2$ islands on top of a SiN/Si chip with multiple holes confirming that $MoS_2$ grains have triangular shapes.

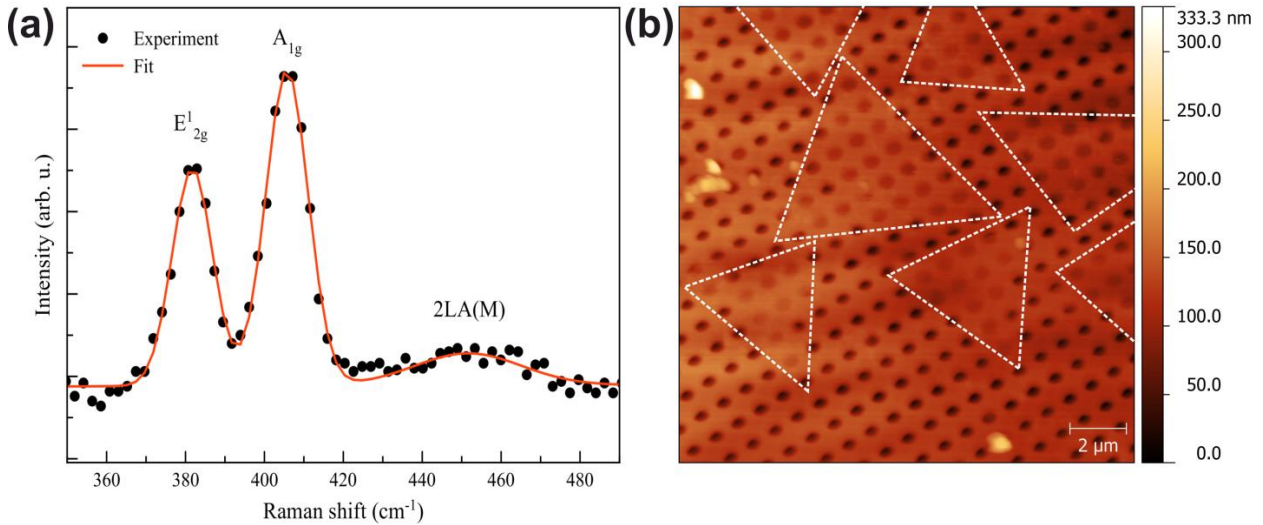

**Supplementary Figure 1.** (a) Raman spectrum of suspended $MoS_2$ showing in-plane $E^1_{2g}$ and out-off plane $A_{1g}$ Raman modes and the 2LA(M) peak measured with a 473 nm laser. (b) AFM topography image of four well-separated $MoS_2$ grains (highlighted with white lines) on a SiN/Si chip with multiple holes.

**Pores between MoS₂ grains**

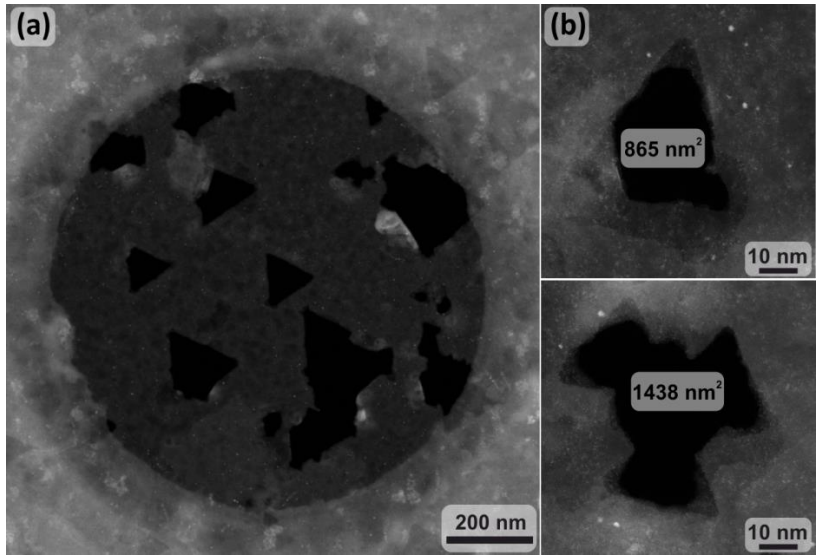

**Supplementary Figure 2.** HAADF-STEM images showing triangular holes between MoS$_2$ grains after prematurely terminated CVD growth.

**Atomic structure of the dislocation core**

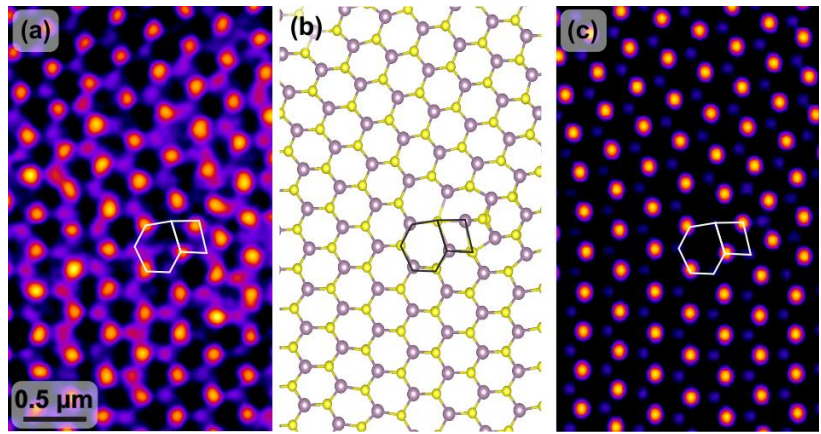

**Supplementary Figure 3.** (a) HAADF-STEM image showing the GB of ML MoS$_2$. (b) An atomistic model of the GB, relaxed by DFT, with a 4|6 dislocation. (c) Simulated HAADF-STEM image of the relaxed model in (b).

**Edge dislocation models**

The strain of an edge dislocation can be described by the Peierls–Nabarro (PN) and the Foreman (FM) models. These models are fitted to our experimental strain distribution around a dislocation core obtained by GPA. The strain of an edge dislocation along the *x* direction is given by PN and FM models as

$$\varepsilon_{xx}^{PN} = -\frac{b}{\pi}\left[\frac{(1-v)y}{4(1-v)^2 x^2 + y^2}\right] \qquad (1)$$

and

$$\varepsilon_{xx}^{FM} = -\frac{b(1-v)}{\pi}\left[\frac{4(1-v)^2 yx^2 + (2a^3 - a^2)y^3}{(4(1-v)^2 x^2 + a^2 y^2)^2}\right]. \qquad (2)$$

Here *b* is the magnitude of Burger vector, *v* the Poisson's ratio (0.25 for ML $MoS_2$)[46], and *a* a fitting parameter. In case of $a = 1$, FM model is equal to PN model.

**Unfiltered version of Figure 5a**

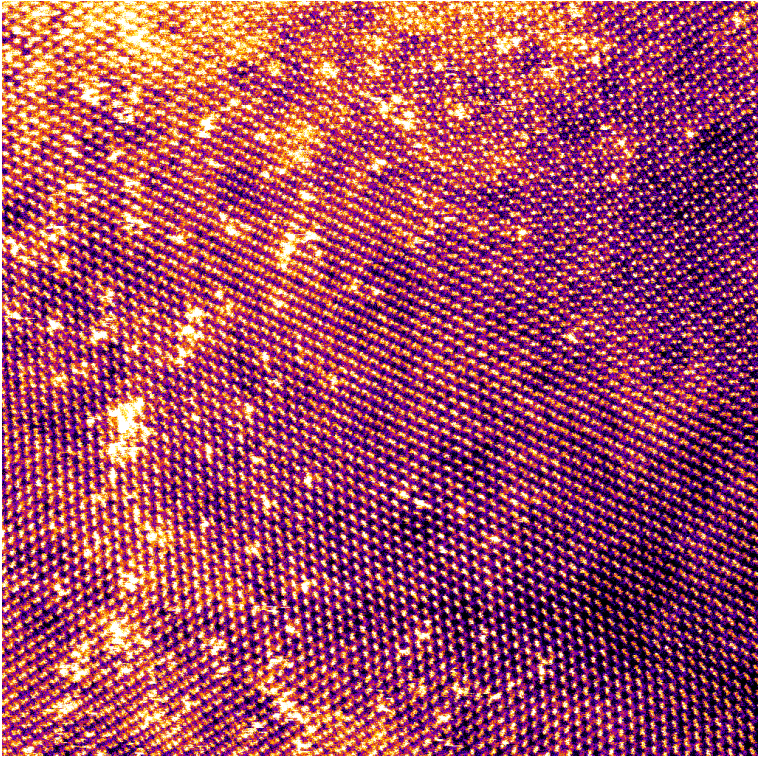

**Supplementary Figure 4.** Unprocessed HAADF-STEM image of bilayer $MoS_2$.

**Analysis of the sample thickness and rippling of bilayers**

In order to confirm the thicknesses of the studied films, we made image simulations for ML and BL $MoS_2$ and compared the intensity profiles of simulated and experimental images, as shown in Supplementary Figure 5. The thickness of layers can also be measured by diffraction experiments (Supplementary Figure 6) because the intensity ratio within the $\{\bar{1}100\}$ family in diffraction patterns of ML and BL $MoS_2$ gives information on layer thicknesses[45]. The intensity ratios in the $\{\bar{1}100\}$ family of ML $MoS_2$ show no tilt or ripple in ML $MoS_2$. In contrast, the intensity ratios for BL $MoS_2$ are in many cases not ~1 like they should be for a flat BL structure [45]. This indicates rippling in one or both of the layers since it can cause such intensity variations[45].

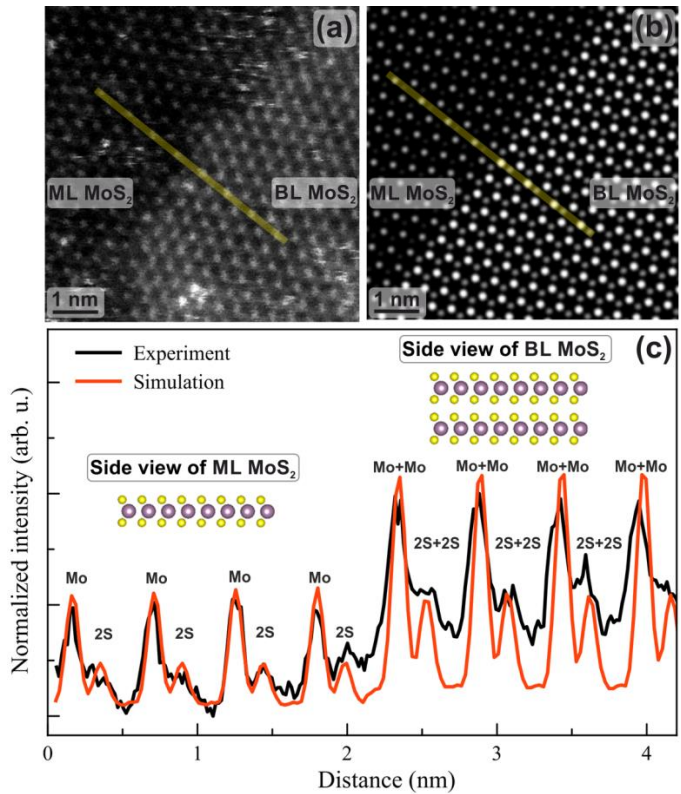

**Supplementary Figure 5.** (a) HAADF-STEM image showing ML and BL $MoS_2$. (b) The corresponding simulated HAADF image. (c) Intensity profiles along the yellow lines on experimental and simulated images.

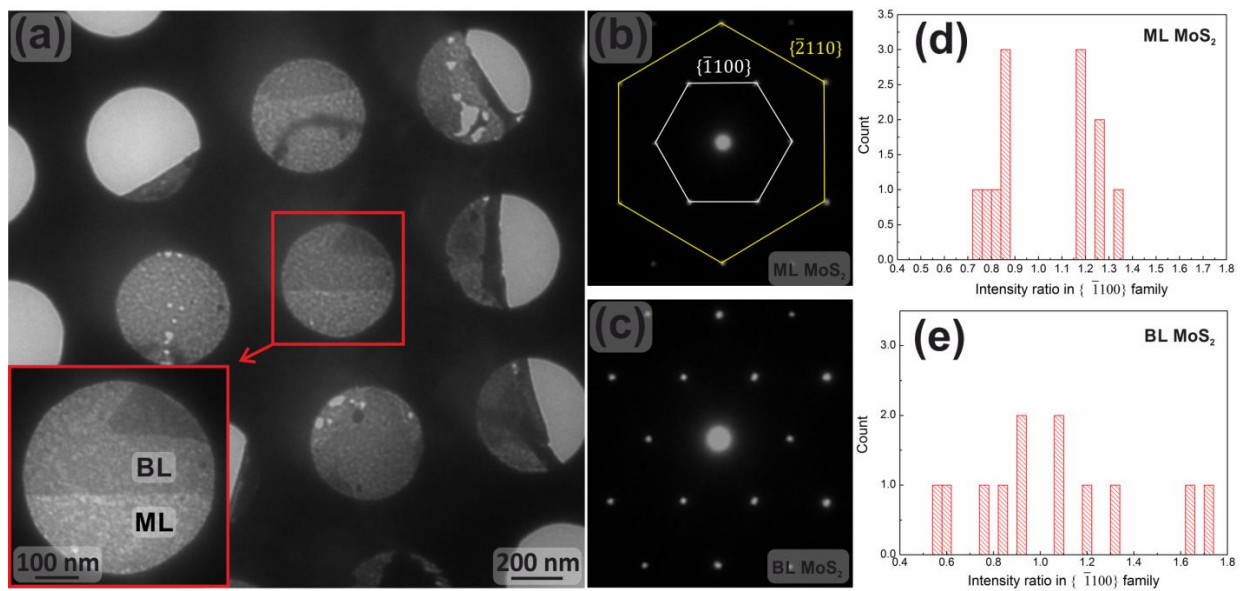

**Supplementary Figure 6.** (a) TEM image of CVD-grown $MoS_2$ on a holey SiN/Si chip. Inset is a close-up image showing ML and BL $MoS_2$ in the red square. (b) and (c) are the diffraction patterns for ML and BL $MoS_2$. (d,e) Histograms indicating the intensity ratio in the $\{\bar{1}100\}$ family for ML and BL $MoS_2$ for several studied sample locations.

**Electron beam-induced nanopores**

The MoS$_2$ GB-NPs shown in the main article have not been created by electron irradiation during the experiment. The formation of NPs under the beam is easily observed during the experiment and is not limited to GBs, as shown in Supplementary Figure 7. Interestingly shear strain maps of beam-induced NPs show that the amount of tensile strain on the sample is reduced during the NP formation while the regions with compressive strain grow. In contrast, the GB-NPs are more tensile strained owing to dislocation cores inducing high levels of strain.

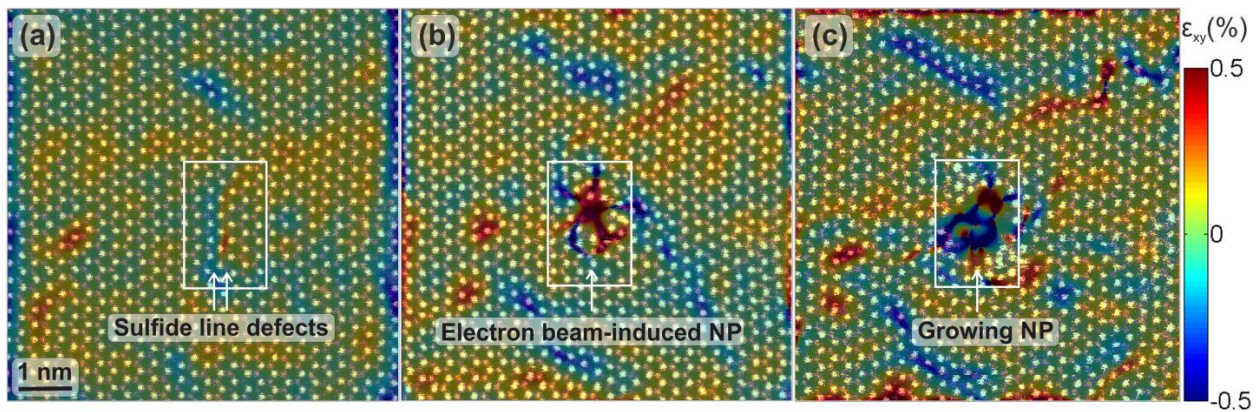

**Supplementary Figure 7.** HAADF-STEM images of MoS$_2$ with a nanopore induced by the electron beam, superimposed with their shear strain maps. (a) Sulfide line defects created during STEM imaging at 60 kV. (b) Formation of the MoS$_2$ NP approximately 30 seconds later. (c) Growing MoS$_2$ NP approximately 50 seconds later. Red and blue correspond to tensile and compressive strain, respectively.

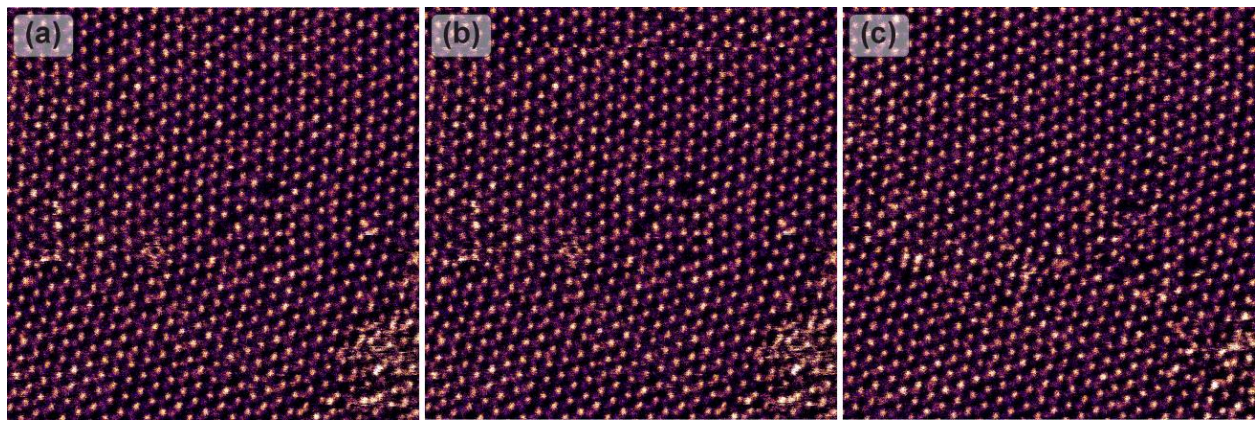

**Supplementary Figure 8**. HAADF-STEM images of MoS$_2$ grain boundary. (a) First scan. (b) 5 seconds later. (c) 50 seconds later.

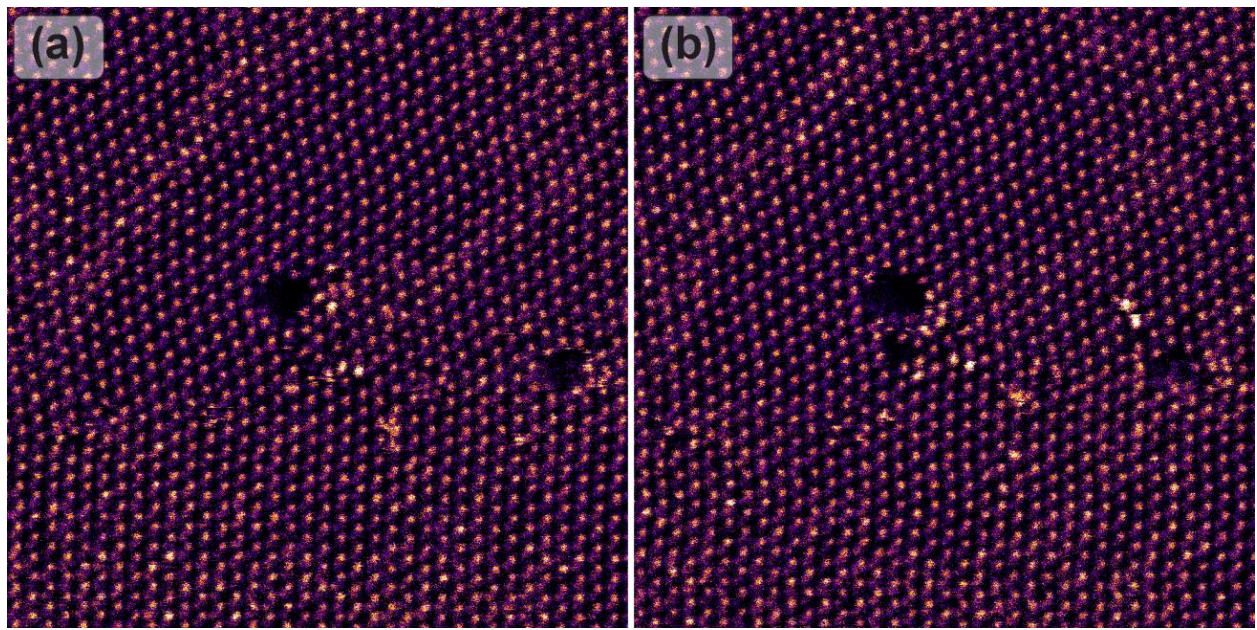

**Supplementary Figure 9**. HAADF-STEM images of MoS$_2$ grain boundary. (a) first scan. (b) 40 seconds later.

**Theoretical results on the bilayer structure in overlap areas**

**Supplementary Table 1.** DFT results for different stacking orders of BL MoS$_2$.

| Stacking | Total energy (eV) | Interlayer distance (Å) | Band gap (eV) |
|---|---|---|---|
| AA | 22.971 | 6.94 | 1.24 |
| AB | 22.926 | 6.52 | 0.99 |
| AB-AA | 22.969 | 7.08 | 1.26 |

**References**

References are listed in the main article.